\documentclass{appolb}
\usepackage{graphicx}

\usepackage{hyperref}
\begin{document}
\title{Central exclusive production of pion pairs\\
in proton-(anti)proton collisions%
\thanks{Presented at EDS Blois 2015: The 16th conference on Elastic and Diffractive scattering,
29 June - 4 July 2015, Borgo, Corsica (France).}%
}
\author{Piotr Lebiedowicz and
Antoni Szczurek\footnote{Also at University of Rzesz\'ow, PL-35-959 Rzesz\'ow, Poland.}
\address{The H. Niewodnicza\'{n}ski Institute of Nuclear Physics
Polish Academy of Sciences,
ul. Radzikowskiego 152, 31-342 Krak\'ow, Poland}
}
\maketitle
\begin{abstract}
We present our results for exclusive
$p p \to p p \pi^+ \pi^-$ or $p \bar p \to p \bar p \pi^+ \pi^-$
processes at high energies.
We discuss the role of new additional absorption corrections in
the non-resonant Lebiedowicz-Szczurek model
mediated by the pomeron and reggeon exchanges.
We discuss also the role of the $\rho^{0}$ and the Drell-S\"oding 
photoproduction mechanism.
We compare our predictions with recent experimental results
obtained by the STAR and CDF Collaborations.
We present predictions for the ALICE, ATLAS and CMS experiments.
Differential distributions in invariant two-pion mass 
and two-dimensional distributions in proton-proton relative azimuthal angle and
transverse momentum of one of the protons are presented.
\end{abstract}
  
\section{Introduction}
There is a growing experimental and theoretical interest 
in understanding soft hadronic processes at high energy; 
for reviews see e.g.~\cite{Albrow:2010yb} and references therein.
One of the reaction which can be relatively easy to measure is 
$pp \to pp \pi^+ \pi^-$ which constitutes 
an irreducible background to three-body processes $p p \to p p M$,
where $M = \rho(770)$, $f_{0}(980)$, $f_{0}(1370)$, $f_{0}(1500)$, $\chi_{c0}$, $f_{2}(1270)$.
There are recently several experimental projects by the COMPASS \cite{Austregesilo:2014oxa}, 
STAR \cite{Adamczyk:2014ofa}, CDF \cite{Albrow_Project_new,Aaltonen:2015uva},
ALICE \cite{Schicker:2014aoa}, ATLAS \cite{Staszewski:2011bg},
CMS \cite{Osterberg:2014mta} and LHCb \cite{Ronan} Collaborations which will measure 
differential cross sections for the $p p \to p p (M \to \pi^{+}\pi^{-})$ reaction(s).
The principal reason for studying central exclusive production of mesons 
is a search for glueballs \cite{Szczurek:2009yk,Ochs:2013gi}.
%states consisting of two or three ``constituent'' gluons,
%There is some evidence from an analysis of the decay modes of the scalar states observed, 
%that the lightest scalar glueball manifests itself through 
%the mixing with nearby $q \bar{q}$ states \cite{Ochs:2013gi,Kirk:2014nwa}.
%The lowest-mass conventional $q \bar{q}$ resonant states have the same quantum numbers. 

The exclusive production of lower mass scalar and pseudoscalar resonances
within a tensor pomeron approach \cite{Ewerz:2013kda}
was examined recently in \cite{Lebiedowicz:2013ika}.
The resonant $\rho^{0}(770) \to \pi^{+}\pi^{-}$ and 
%non-resonant (Drell-S\"oding)
%photon-pomeron/reggeon production 
%We consider also the exclusive central production of $\rho(770)$ resonance
nonresonant background (Drell-S\"oding) mechanism via the photon-pomeron/reggeon exchanges
was considered in \cite{Lebiedowicz:2014bea}.
%in the context of theoretical concept of tensor pomeron proposed in Ref.\cite{Ewerz:2013kda}. 
%Due to its quantum numbers this resonance state can be
%produced only by photon-pomeron/reggeon or reggeon-pomeron exchanges.
%The details have been presented in \cite{Lebiedowicz:2014bea}
There the effective vertices and propagators 
have been taken from Refs.~\cite{Ewerz:2013kda} and \cite{Bolz:2014mya}.
The coupling parameters of tensor Regge exchanges
are fixed based on the HERA data for the $\gamma p \to \rho^{0} p$ reaction.
%In the amplitude for the $\gamma p \to \rho^{0} p$ subprocess
%we included both pomeron and $f_{2 \Reg}$ exchanges.

Some time ago two of us proposed a simple phenomenological Regge-like model 
for the $\pi^{+} \pi^{-}$-continuum mechanism \cite{Lebiedowicz:2009pj}.
For early studies of two pion production, see Refs.~\cite{Pumplin:1976dm,Desai:1978rh}.
Recently, the Lebiedowicz-Szczurek model \cite{Lebiedowicz:2009pj}
was implemented in GenEx MC \cite{Kycia:2014hea}.
Another related generator is DIME MC \cite{Harland-Lang:2013dia}.
In Refs.~\cite{Lebiedowicz:2011nb,HarlandLang:2012qz} 
the continuum background was considered to the 
production of $\chi_c(0^{+})$ decaying into $\pi^{+} \pi^{-}$ or $K^{+} K^{-}$ channels.
%There is naturally mixing between the gluonic and $q\bar{q}$ systems.
For exclusive production of other mesons
see e.g. \cite{Lebiedowicz:2010yb,Cisek:2011vt,Lebiedowicz:2013vya},
where mainly the non-central processes were discussed.

\section{Sketch of formalism}

\begin{figure}[htb]
\centerline{%
(a)\includegraphics[width=4cm]{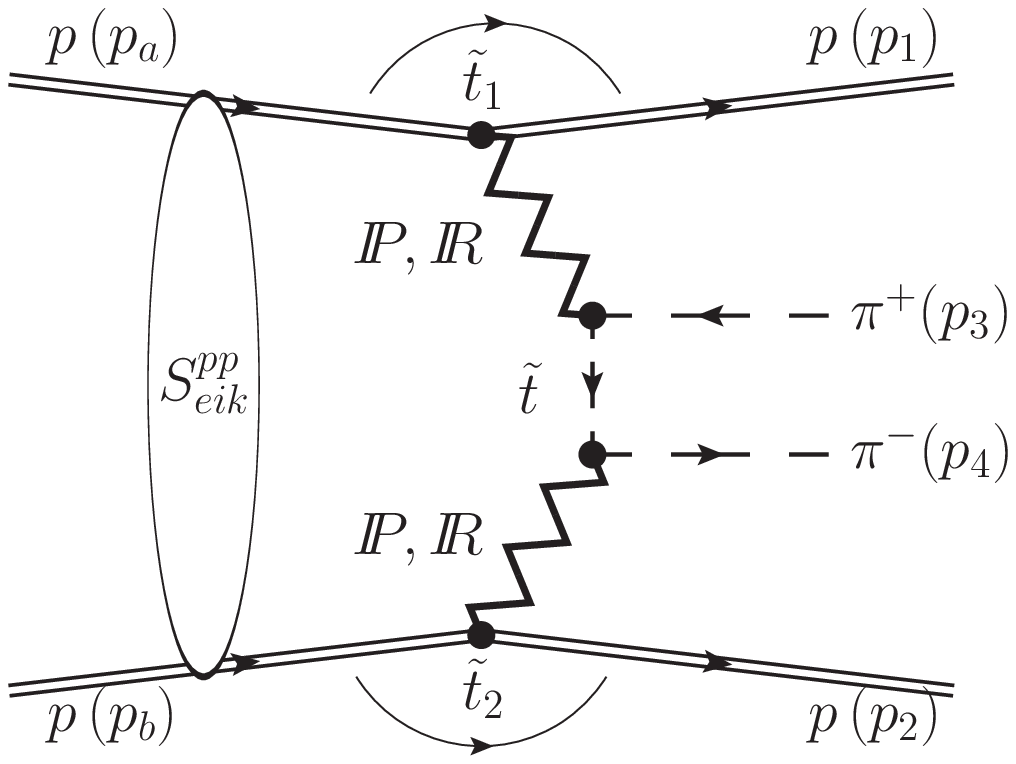}
(b)\includegraphics[width=4cm]{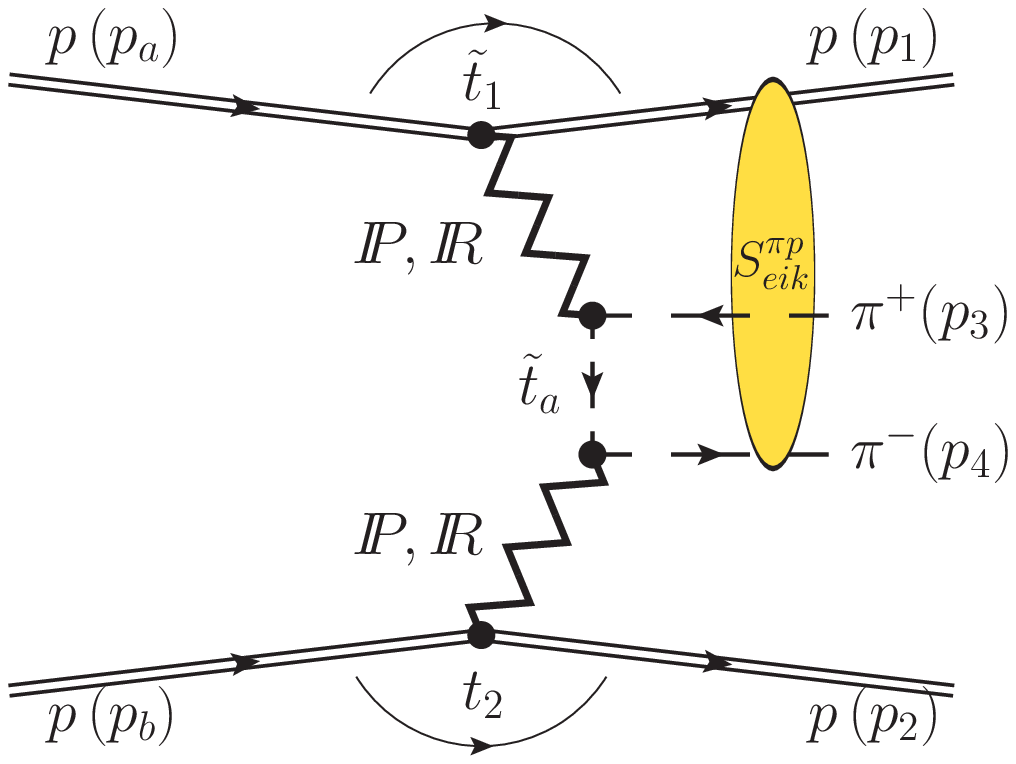}}
%}
\caption{The diagrams for double pomeron/reggeon
central exclusive continuum $\pi^+ \pi^-$ production including 
the absorptive corrections due to the $pp$ interaction (diagram (a))
and due to the $\pi p$ interaction (diagram (b)).}
\label{Fig:1}
\end{figure}

The Born amplitude with the intermediate $\pi$-exchange can be written
\begin{eqnarray} 
\mathcal{M}=
M_{13}(s_{13},t_1)
\frac{F_{\pi}^{2}(t)}{t-m_{\pi}^{2}}%F_{\pi}(t)
M_{24}(s_{24},t_2)
+
M_{14}(s_{14},t_1)
\frac{F_{\pi}^{2}(u)}{u-m_{\pi}^{2}}%F_{\pi}(u)
M_{23}(s_{23},t_2),
\label{Regge_amplitude}
\end{eqnarray}
where the subsystem amplitudes $M_{ij}(s_{ij},t_{i})$ 
denote ``interaction'' between forward proton ($i=1$)
or backward proton ($i=2$) and one of the two pions
($j=3$ for $\pi^{+}$ or $j=4$ for $\pi^{-}$). 
In the Lebiedowicz-Szczurek model \cite{Lebiedowicz:2009pj,Lebiedowicz:2011nb,Lebiedowicz:thesis}
the parameters of pomeron and subleading reggeon exchanges
were adjusted to describe total and elastic $\pi N$ scattering.
The largest uncertainties in the model are due
to the unknown form of off-shell pion form factor $F_{\pi}(k^{2})$ 
and the absorption corrections
calculated here in the eikonal approximation, see diagrmas in Fig.~\ref{Fig:1}.
Recently, in \cite{Lebiedowicz:2015eka} we estimated new absorptive corrections
due to the proton-pion rescattering in the final state.
%and discussed their influence 
%on distributions in different kinematic variables.
%--------------------------------------------------------------------
%--------------------------------------------------------------------

\section{Results and Conclusions}

\begin{figure}
[htb]
\centerline{%
\includegraphics[width=5.cm]{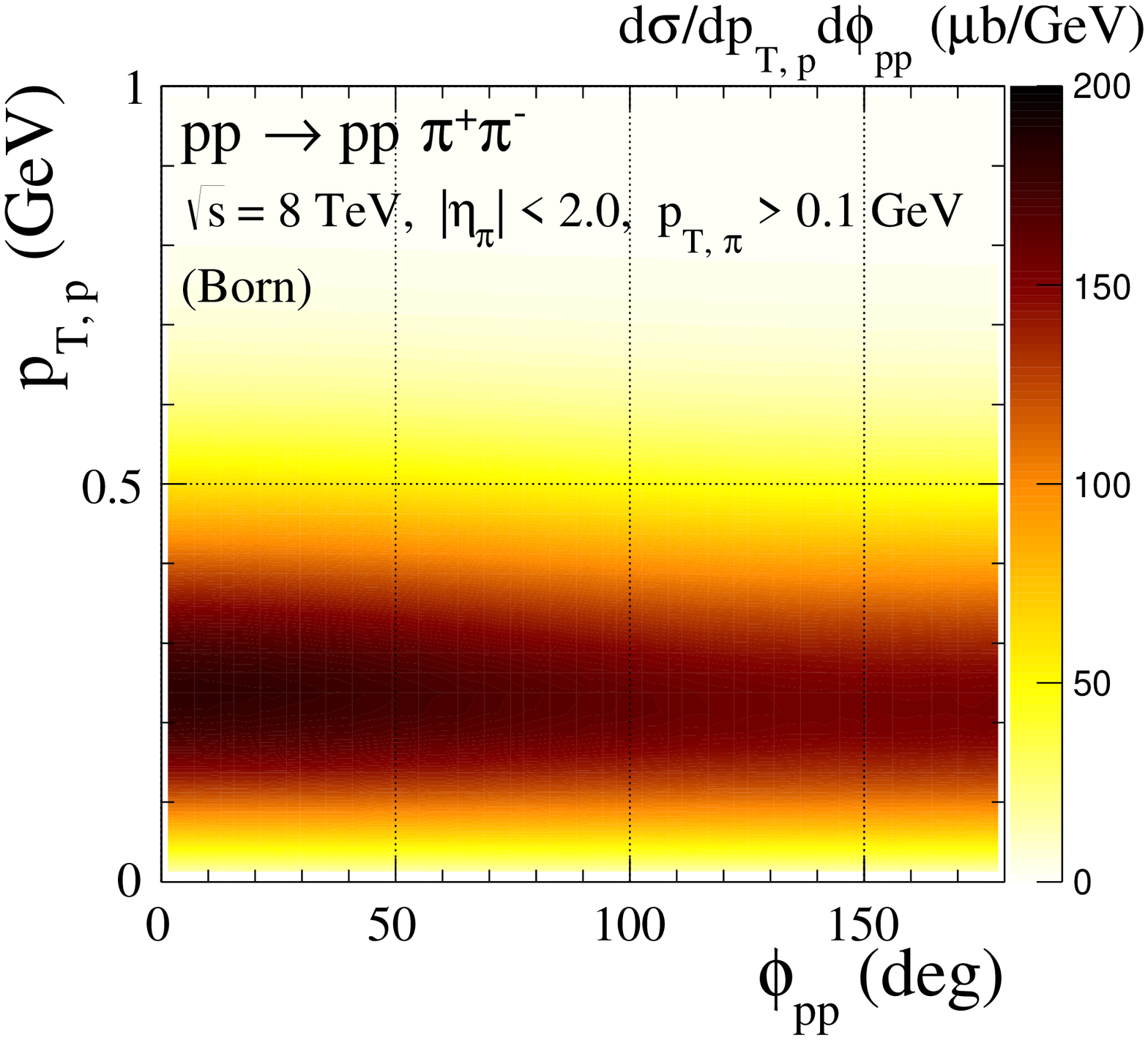}
\includegraphics[width=5.cm]{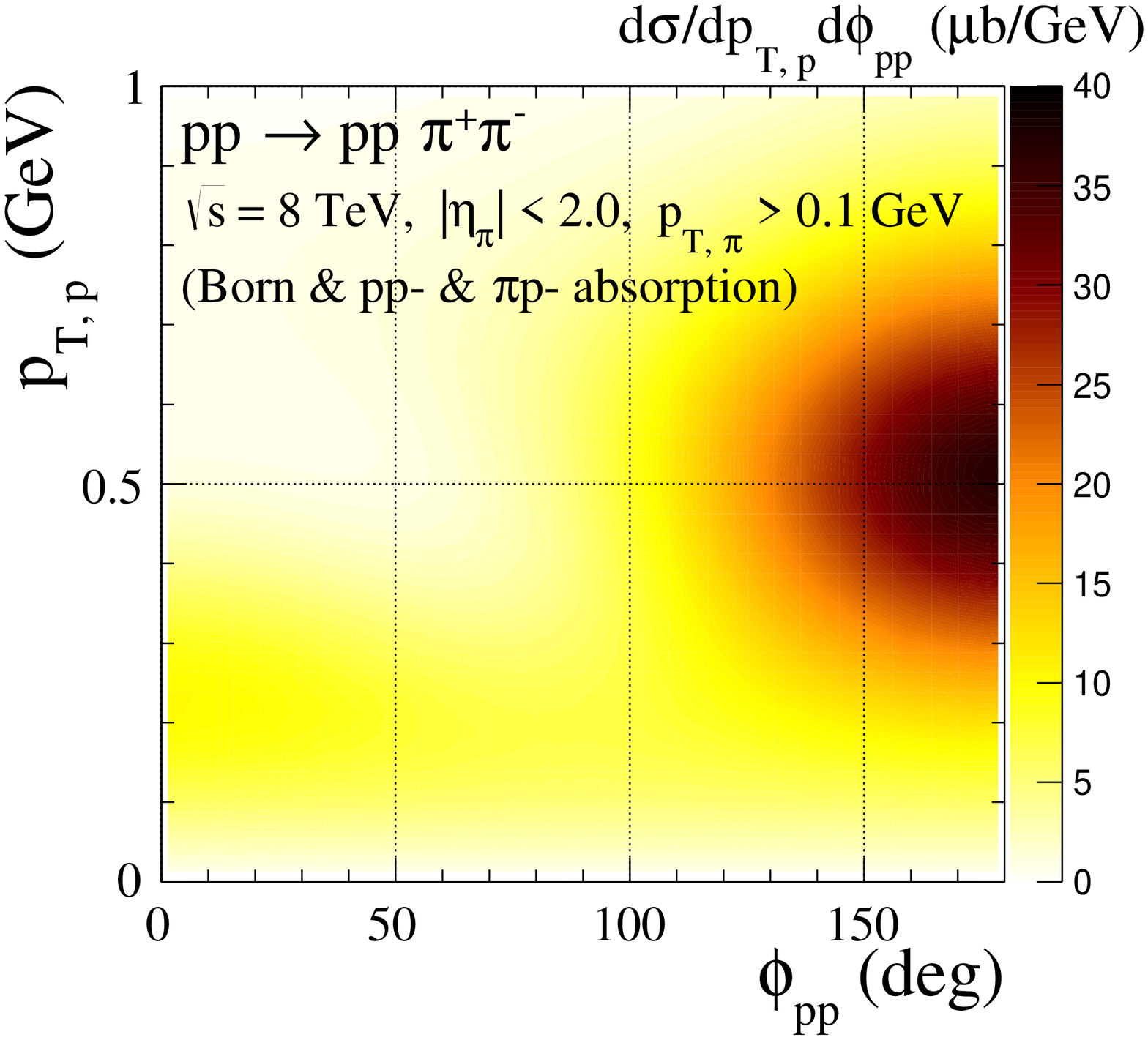}}
%}
\caption{Two dimensional distributions in $p_{t,p}$ and $\phi_{pp}$
at $\sqrt{s} = 8$~TeV with the CMS kinematical cuts.
We show the distributions without and with the absorption corrections, see Fig.~\ref{Fig:1}.
In this calculation we have used 
%the cut-off parameter 
$\Lambda_{off,E} = 1.6$~GeV.}
\label{Fig:2}
\end{figure}

Recently, we have discussed the role of soft $pp$- and $\pi p$-rescattering corrections.
In Fig.~\ref{Fig:2} we show two-dimensional
distributions in proton-proton relative azimuthal angle and
transverse momentum of one of the protons
without (the left panel) and with (the right panel) the absorption corrections.
The absorption effects lead to substantial damping of the cross section
and to a shape deformation of differential distributions 
in contrast to the commonly used uniform factor known as gap survival factor.
The damping depends on the collision energy and kinematical variables.
The ratio of full and Born cross sections 
$\langle S^{2}\rangle$ (the gap survival factor)
is approximately 0.20 (STAR), 0.09 (CDF), 0.12 (LHC).
This could be verified in a future in experiments when both protons 
are measured, such as ATLAS-ALFA or CMS-TOTEM.
%--------------------------------------------------------------------
%--------------------------------------------------------------------

\begin{figure}[htb]
\centerline{%
\includegraphics[width=6.cm]{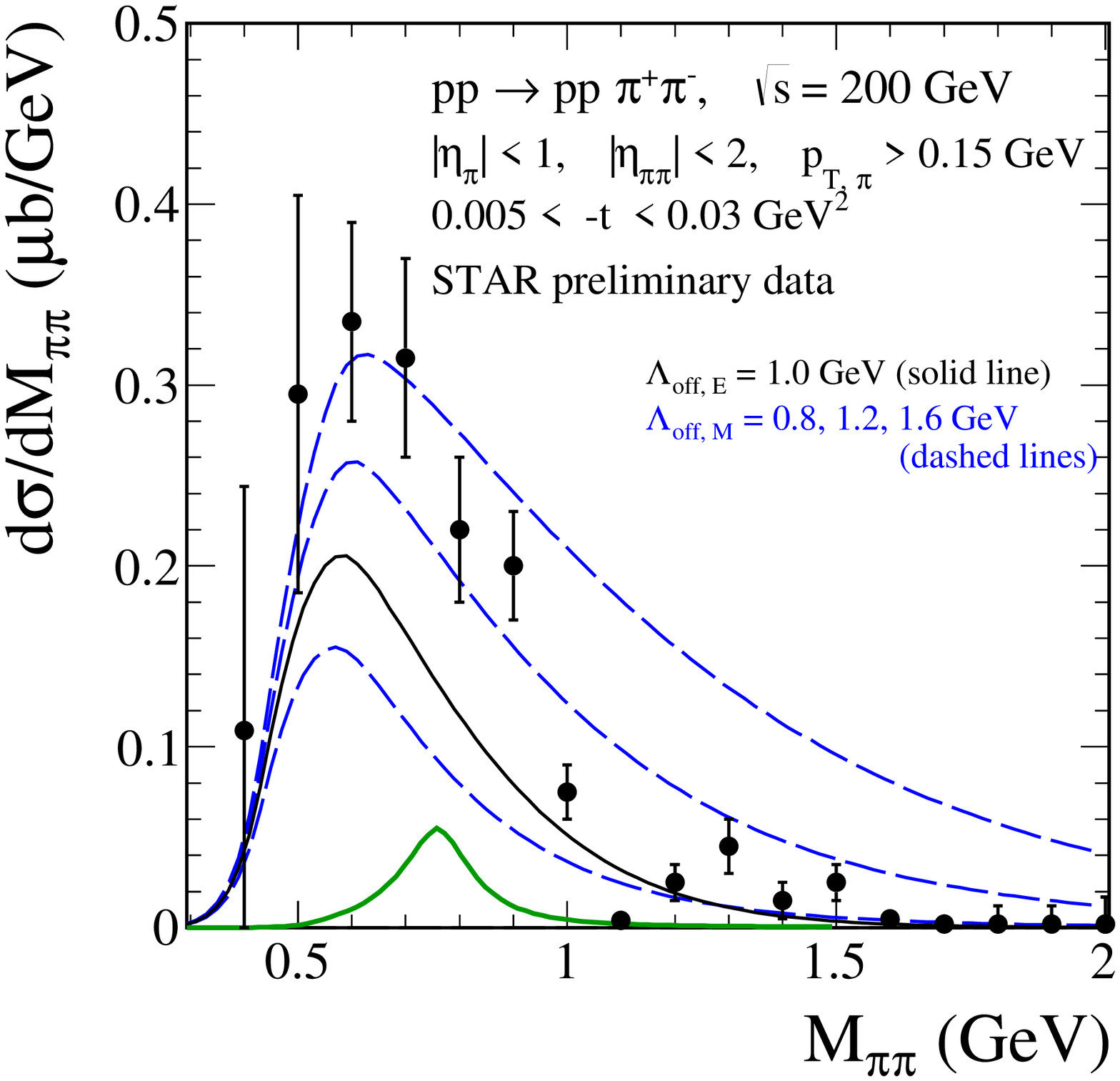}
\includegraphics[width=6.cm]{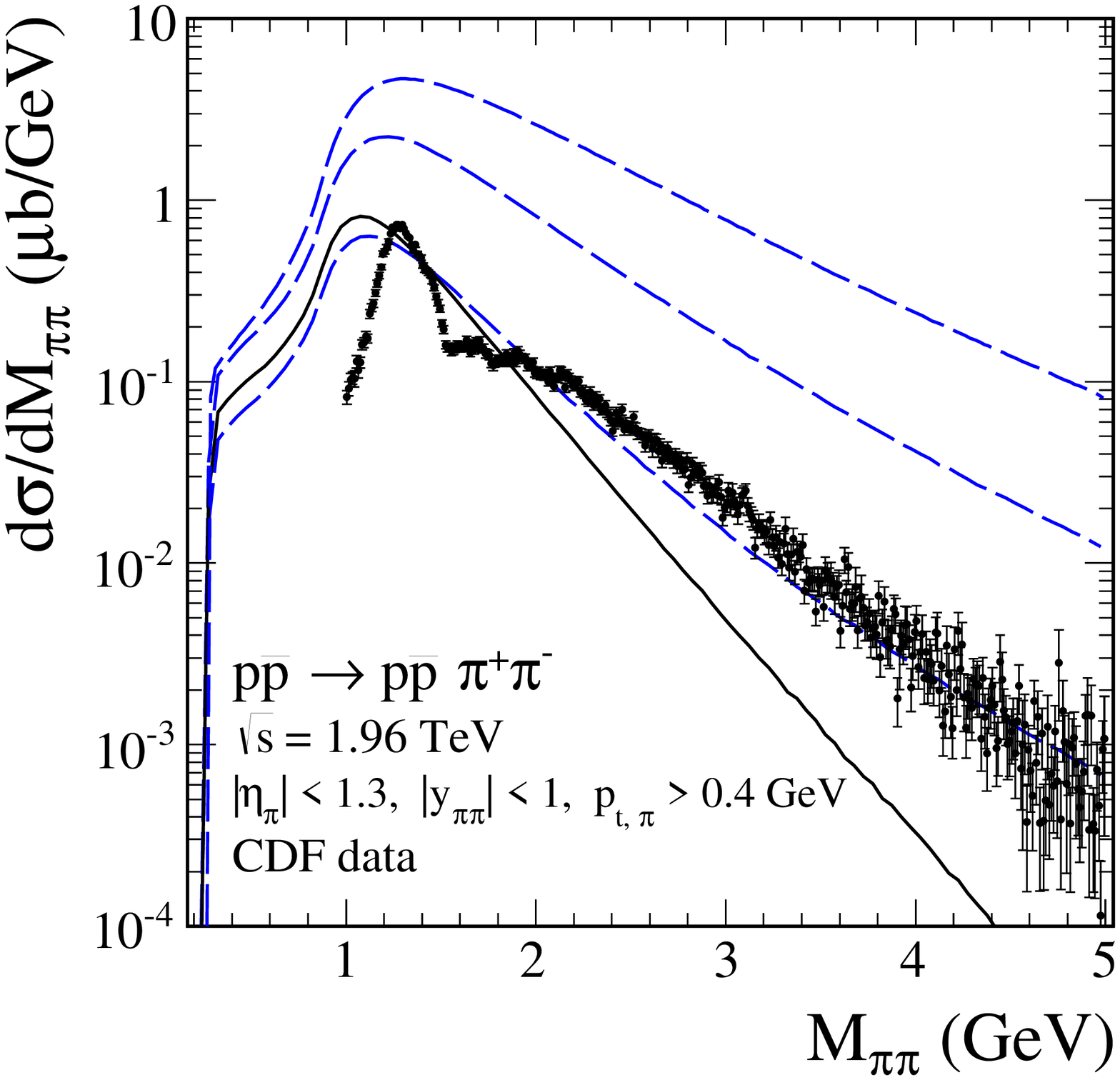}}
\centerline{%
\includegraphics[width=6.cm]{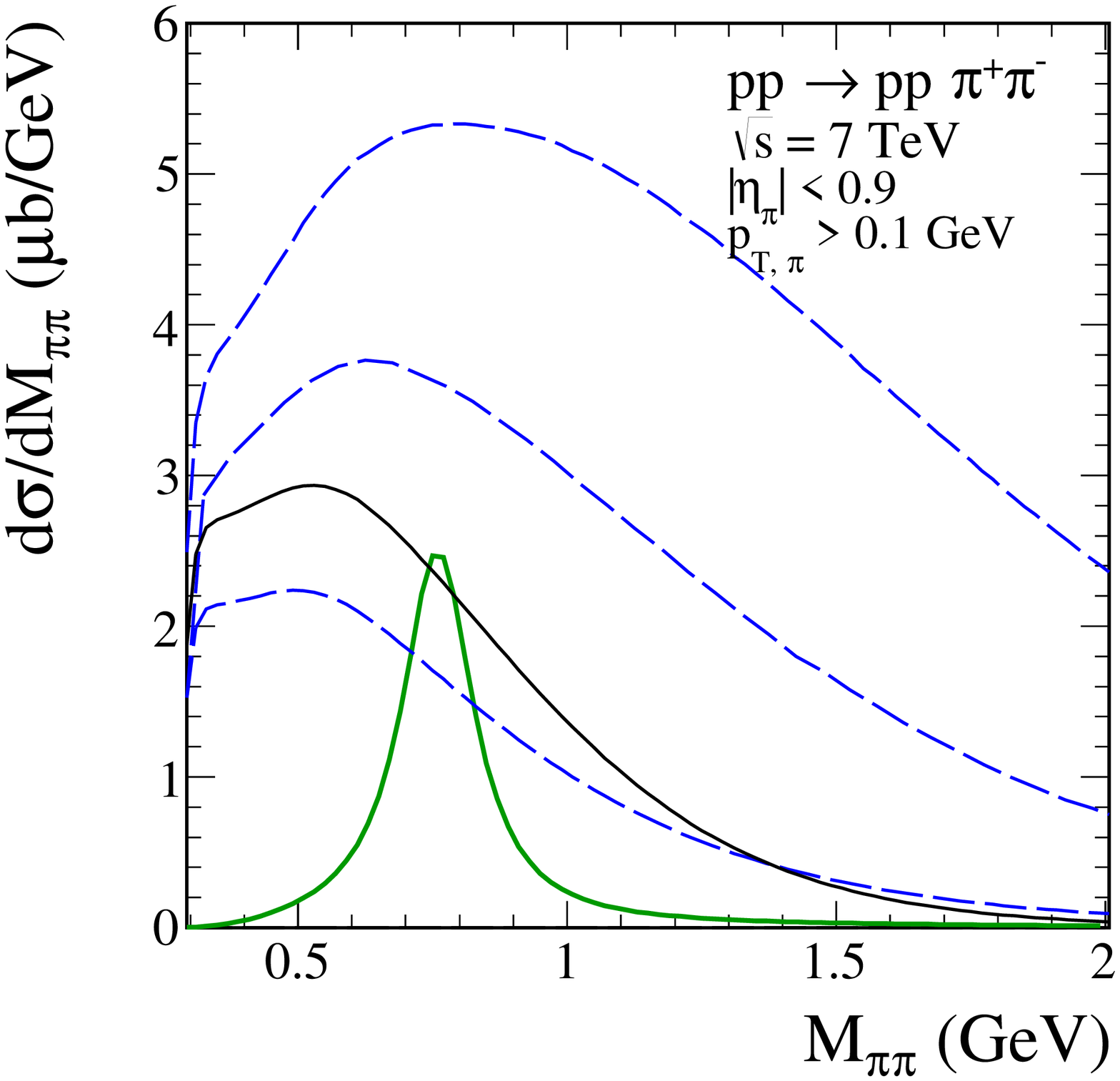}
\includegraphics[width=6.cm]{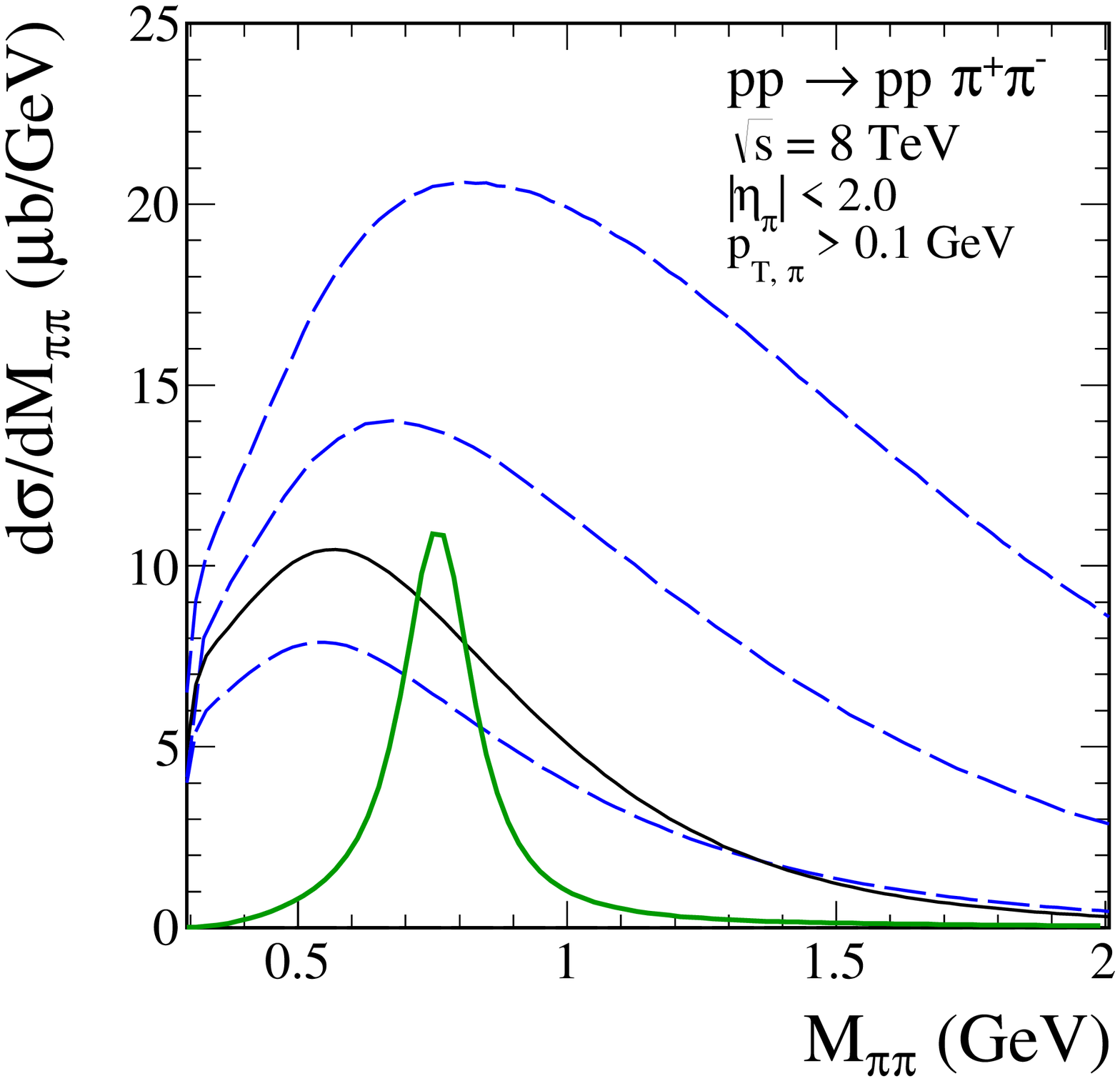}}
%}
\caption{Two-pion invariant mass distribution at $\sqrt{s}=0.2, 1,96, 7$ and 8~TeV
with the experimental kinematical cuts indicated in the legend.
We show results for the double pomeron/reggeon contribution
including all absorption corrections discussed in \cite{Lebiedowicz:2015eka}.
The blue dashed lines represent the results obtained 
for the monopole form factors in Eq.~(\ref{Regge_amplitude})
$F_{\pi}(k^{2})=(\Lambda^{2}_{off,M} - m_{\pi}^{2})/(\Lambda^{2}_{off,M} - k^{2})$
and $\Lambda_{off,M} = 0.8, 1.2, 1.6$~GeV (from bottom to top)
while the black solid lines are for the exponential form 
$F_{\pi}(k^{2})=\exp[(k^{2}-m_{\pi}^{2})/\Lambda^{2}_{off,E}]$ 
and $\Lambda_{off,E} = 1.0$~GeV.
The green solid lines represent results for the photoproduction contribution \cite{Lebiedowicz:2014bea}.
The STAR \cite{Adamczyk:2014ofa} and CDF \cite{Aaltonen:2015uva,Albrow_Project_new}
data are shown for comparison.}
\label{Fig:3}
\end{figure}

Now we wish to compare predictions of the Lebiedowicz-Szczurek model
with full absorption
with the recent STAR and CDF data and present predictions for the LHC experiments.
In Fig.~\ref{Fig:3} we show two-pion invariant mass distribution
for different experiments with relevant kinematical cuts.
At $M_{\pi\pi} \simeq 1$~GeV the STAR and CDF data show a minimum 
due to interference of the $f_{0}(980)$ resonance contribution
with the non-resonant background contribution.
At higher $M_{\pi\pi}$ some structures
could be attributed most probably to $f_{2}(1270)$, $f_{0}(1370)$, $f_{0}(1500)$, and $f_{0}(1710)$
resonant states
\footnote{The $f_{0}(1500)$ and the $f_{0}(1710)$ mesons 
are considered to be scalar glueball candidates \cite{Ochs:2013gi}, 
but mixing with quarkonium states complicates the issue.}.
One can observe that our predictions are quite sensitive 
to the form of the off-shell pion form factors
%(\ref{off-shell_form_factors_exp}) or (\ref{off-shell_form_factors_mon})
and depend on the value of the cut-off parameters $\Lambda_{off}$.
%and for the different values of the off-shell-pion form factor parameters.
If we describe the maximum of the cross section around $M_{\pi \pi} \sim$~0.6 GeV 
measured by the STAR experiment we overestimate the cross section
in the interval 1~$< M_{\pi \pi} <$~2~GeV. 
A part of the effect may be related to
an enhancement of the cross section due to $\pi \pi$ low-energy final state interaction \cite{Pumplin:1976dm,Au:1986vs,Lebiedowicz:2009pj}.
Therefore, we might expected that at higher masses the non-resonant model 
gives realistic predictions
with the off-shell pion form factor parameter $\Lambda_{off} \approx 1$~GeV.
%--------------------------------------------------------------------
%--------------------------------------------------------------------

Using the tensor-pomeron approach we have included
the $\rho(770)$ resonance and the non-resonant Drell-S\"oding contributions
(the green solid lines in Fig.~\ref{Fig:3})
which constitute the main source of $P$-wave in the $\pi^+ \pi^-$ channel 
in contrast to even waves populated in double-pomeron/reggeon processes.
Due to the photon propagators occurring in these diagrams
we expect these processes to be most important
when at least one of the protons is undergoing only 
a very small momentum transfer $|t|$.
We have observed that at midrapidities, imposing e.g. a cut $|\eta_{\pi}|<0.9$,
the photoproduction term could be visible in experiments.
The absorptive corrections for photon induced reactions
lead to only about 10\% reduction of the cross section.

It would clearly be interesting to extend the studies of central meson production
for other resonances such as the $f_0(980)$ and $f_2(1270)$ mesons
decaying into $\pi^{+} \pi^{-}$ channel.
Then the interference effects of the resonance signals 
with the two-pion continuum has to be included in addition. 
This requires a consistent model of the resonances and the non-resonant background.
The interference effects may depend on $t_{1}$ and $t_{2}$ 
that are very different for RHIC, Tevatron and LHC experiments.
This aspects should be addressed in a future \cite{LNS}.

\textbf{Acknowledgements}
We would like to thank Otto Nachtmann for collaboration on the photoproduction contribution.
This research was partially supported by the MNiSW Grant No. IP2014~025173 (Iuventus Plus),
the START fellowship from the Foundation for Polish Science,
and by the Polish National Science Centre Grant No. DEC-2014/15/B/ST2/02528 (OPUS)
as well as by the Centre for Innovation and Transfer of Natural Sciences 
and Engineering Knowledge in Rzesz\'ow.

%------------------------------------------------------------------
%\include{bibliography}
%------------------------------------------------------------------

\end{document}